\documentclass[aps,preprint,groupedaddress,nofootinbib,longbibliography]{revtex4-1}

\usepackage{amssymb}
\usepackage[breaklinks,colorlinks,citecolor=green,urlcolor=blue,linkcolor=red]{hyperref}
\usepackage{graphicx}
\usepackage{epstopdf}
\usepackage{diagbox}
\usepackage{float}
\usepackage{subfigure}
\usepackage{multirow}

\begin{document}

%\preprint{}

\title{\boldmath Data-based analysis of the Forward-Backward Asymmetry in $B^\pm \to K^\pm K^\mp K^\pm$}


\author{Rui Hu}
\author{Zhen-Hua Zhang}
\email{zhangzh@usc.edu.cn}
%\thanks{}
\affiliation{School of Nuclear Science and Technology, University of South China, Hengyang, 421001, Hunan, China}


\date{\today}

\begin{abstract}
An analysis of the Forward-Backward Asymmetry (FBA) in the decay $B^\pm \to K^\pm K^\mp K^\pm$ is carried out based on the LHCb data.
It is found that the large FBA observed for the invariant mass of the $K^+ K^-$ pair around 1.5 GeV can be explained by the interference of the amplitudes between the resonances with even and odd spins, where the former can be the spin-0 $f_0(1500)$ resonance plus a non-resonance $s$-wave, while the latter is a spin-1 resonance which is probably $\rho^0 (1450)$.
This is in contradiction with the conclusion of former experimental analysis such as one of BaBar's in Phys. Rev. D 85, 112010, according to which the analysis showed no signal of spin-odd resonances at all when the invariant mass of the $K^+ K^-$ pair around 1.5 GeV.
While according to the analysis of the current paper, the existence of the spin-odd resonances such as $\rho^0(1450)$ is inevitable for the explanation of the large FBA in this region.
The analysis also shows that the $C\!P$ asymmetry of the decay channel $B^\pm\to \rho^0(1450) K^\pm$ is about $(-3.4\pm3.0)\%$.
We suggest our experimental colleagues to perform a closer analysis to this channel.
We also suggest to perform the measurements of the FBAs (as well as the FB-$C\!P$As) in other three-body decay channels of beauty and charmed mesons, as it is helpful for resonance analysis.
\end{abstract}

%\pacs{}
%\keywords{CP violation, heavy baryon,}

\maketitle

%\section{}
%\subsection{}
%\subsubsection{}

\section{Introduction\label{se:intro}}

The observables Forward-Backward Asymmetries (FBAs) have played an important role in the history of particle physics.
Examples include the discovery of the parity violation of weak interaction \cite{Lee:1956qn,Wu:1957my,Garwin:1957hc}, the precision measurement of the $Z$ boson \cite{ALEPH:2005ab}, the study of the lepton universality \cite{Ali:1991is,Alok:2009tz,Belle:2009zue}.
The introduction of the FBA and the FBA induced $C\!P$ Asymmetry (FB-$C\!P$A) to the hadronic multi-body decays of beauty and charmed mesons provides a good approach for isolating the interfering effects between near-by resonances \cite{Zhang:2021zhr}, which is helpful for the understanding of the behaviour of the $C\!P$ violation, the resonance spectroscopy, as well as the low energy  quantum chromodynamics (QCD).

The decays of $B$ meson are excellent probes for New Physics (NP) indirectly via the study of $C\!P$ violation ($C\!P$V) and rare decays, as well as good places for improving our understanding of QCD at low energy via spectroscopy study of resonances, among which the hadronic multi-body decays of $B$ mesons becomes increasingly important.
For the former case, the lepton universality in decays $B\to K^{(\ast)}l^+l^-$ has gained a lot of attention form both the theoretical and experimental sides \cite{LHCb:2017avl,Belle:2019oag,LHCb:2019hip,Bordone:2016gaq,Bobeth:2007dw,Geng:2017svp,Alonso:2014csa,DAmico:2017mtc}.
For the latter, QCD exotic states such as the pentaquark states were also first observed in hadronic multi-body decays of $B$ meson \cite{Belle:2003nnu,LHCb:2015yax,LHCb:2016axx,LHCb:2019kea}.

This three-body $B$ meson decay channel $B^\pm\to K^\pm K^\mp K^\pm$ has been studies experimentally by BaBar \cite{BaBar:2006hyf,BaBar:2012iuj}, Belle\cite{Belle:2004drb}, and LHCb \cite{LHCb:2014mir}, in which a structure referred as $f_X(1500)$ when the invariant mass of one $K^+ K^-$ pair is around $1.5$ GeV was reported by BaBar and Belle.
Although it can be explained by $f_0(1500)$ or a combination of some even-spin resonances such as $f_0(1500)$ and $f_2'(1525)$ for the BaBar and the Belle cases, the nature of $f_X(1500)$ is still unclear.
Recent theoretical investigations via perturbative QCD approach indicates that the $f_X(1500)$ structure is probably the spin-1 resonance $\rho^0(1450)$ \cite{Zou:2020mul,Zou:2020atb,Liu:2021sdw}.

The LHCb data in Ref. \cite{LHCb:2014mir} provide us the opportunity to investigate the nature of $f_X(1500)$ via the FBA and the FB-$C\!P$A in the decay channel $B^\pm\to K^\pm K^\mp K^\pm$.
The evident large FBA when the invariant mass of $K^+ K^-$ pair is around $1.5$ GeV, which can even be clearly seen from the events distributions in the Dalitz plots of this channel, implies strongly the presence of a spin-odd resonance, which could probably be $\rho^0(1450)$, with reasons which will be explained in this paper.
This is clearly in contradiction with the former analysis performed by BaBar and Belle.

The remainder of this paper is structured as follows.
In Sec. \ref{sec:IntroFBA}, we present the definition of the FBA and the FB-$C\!P$A for the three-body decays of $B$ meson, followed with a brief discussion.
In Sec. \ref{sec:AnalysisFBA}, we perform the analysis of the FBA and the FB-$C\!P$A of $B^\pm\to K^\pm K^\mp K^\pm$ based on the data of LHCb in Ref. \cite{LHCb:2014mir}, according to which it is found that the large FBA strongly indicates the presence of the $p$-wave resonances $\rho^0(1450)$.
In the last section, we make our conclusion.

\section{\boldmath The FBA and the FB-$C\!P$A in three-body decays of heavy meson \label{sec:IntroFBA}}
\begin{figure}[h]
	\centering
			 \includegraphics[width=.6\linewidth]{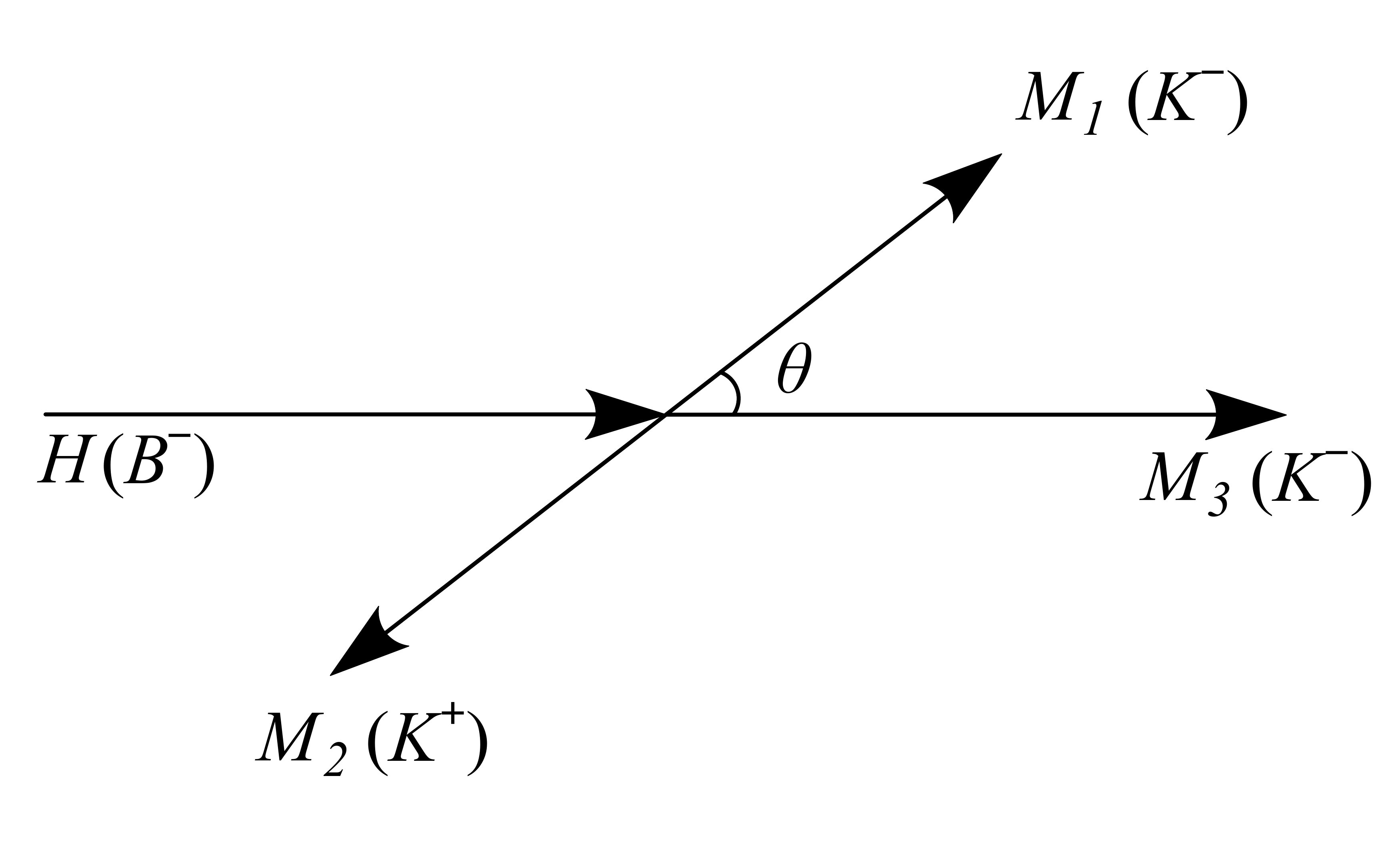}
			\caption{The definition of $\theta$ in the c.m. frame of the $M_1M_2$ system.}\label{fig:def-theta}
		% \label{level}
\end{figure}

For a three-body decay process of a beauty or a charmed meson, $H\to M_1 M_2 M_3$, with $M_j$ ($j=1,2,3$) being three pseudo-scalar mesons which can only decay via electro-weak interactions, we denote the angle between the momentum of $M_1$ and that of the initial particle $H$ in the c.m. frame of the $M_1M_2$ system as $\theta$ (see FIG. \ref{fig:def-theta} for illustration), and the invariant mass squared of $M_iM_j$ system as $m_{ij}^2$.
One has the relation
\begin{equation}%\label{}
\cos\theta\equiv\frac{\vec{p}_1^\ast\cdot \vec{p}_3^\ast}{|\vec{p}_1^\ast||\vec{p}_3^\ast|}=\frac{m^2_{23}-(m^2_{{23},\text{max}}+m^2_{{23},\text{min}})/2}{(m^2_{{23},\text{max}}-m^2_{{23},\text{min}})/2},
\end{equation}
where $\vec{p}_j^\ast$ is the momentum of $M_j$ in the c.m. frame of the $M_1M_2$ system, $m^2_{{23},\text{max (min)}}$ is the maximum (minimum) value of $m^2_{23}$ constrained by the phase space.
The Forward-Backward Asymmetry (FBA), which describes the preference of the flying direction of $M_1$ with respect to that of $H$ in the c.m. frame of the $M_1M_2$ system, is defined as \cite{Zhang:2021zhr}
\begin{equation}\label{eq:defFBA}
  A^{FB}=\frac{\int_0^1\langle\left|\mathcal{A}\right|^2\rangle d\cos\theta-\int_{-1}^0\langle\left|\mathcal{A}\right|^2\rangle d\cos\theta} {\int_{-1}^1\langle \left|\mathcal{A}\right|^2\rangle d\cos\theta},
\end{equation}
where the notion ``$\langle \cdots\rangle$'' represents integration over the invariant mass squared $m^2_{12}$, $\langle \left|\mathcal{A}\right|^2\rangle\equiv\int_a^b \frac{(m^2_{{23},\text{max}}-m^2_{{23},\text{min}})}{2}\left|\mathcal{A}\right|^2 dm^2_{12}$, with $[a,b]$ the interval that the integration was performed on.
By expressing the decay amplitudes in terms of partial waves,
\begin{equation}\label{eq:PWamp}
  \mathcal{A}=\sum_l a_{l}P_{l}(\cos\theta),
\end{equation}
one find that the FBA depends on the interferences of even and odd partial waves:
\begin{equation}\label{eq:AFB}
  A^{FB}= \frac{2}{\sum_{j} \left[{\langle \left|a_j\right|^2\rangle}/{(2j+1)}\right]}\sum_{{\text{even~}} l \atop {\text{odd~}} k} f_{lk}\Re\left(\langle a_{l}a_{k}^{\ast}\rangle\right),
\end{equation}
where $f_{lk}\equiv\int_0^1 P_{l}P_{k}d\cos\theta=\frac{(-)^{(l+k+1)/2}l!k!}{2^{l+k-1}(l-k)(l+k+1)[(l/2)!]^2\{[(k-1)/2]!\}^2}$ \cite{Byerly}.
From this equation, one can see that the numerator contains {\it only} the interference term between even- and odd-waves.
This implies that large FBA around even (odd)-wave resonances usually indicates the interference with nearby odd (even)-wave contributions
\footnote{
Since $M_1$ and $M_2$ are spin-0 particles, the spin of the resonance decaying into them equals to the angular momentum between them.}.
It is impossible to generate a large FBA with only the presence of even or odd waves.

The Forward-Backward Asymmetry induced $C\!P$A (FB-$C\!P$A) is defined as the difference between FBAs of the pair of $C\!P$-conjugate processes, which reads
\begin{equation}\label{eq:ACPFB}
  A_{CP}^{FB}\equiv\frac{1}{2}\left( A^{FB}- \overline{A}^{FB}\right),
\end{equation}
where $ \overline{A}^{FB}$ is the FBA of the $C\!P$-conjugate process $\overline{H}\to \overline{M}_1 \overline{M}_2 \overline{M}_3$, the factor $1/2$ is introduced so as to make sure the value of the $A_{CP}^{FB}$ lies between $-1$ and 1.
One immediately see from Eqs. (\ref{eq:AFB}) and (\ref{eq:ACPFB}) that the FB-$C\!P$A has the ability of isolating $C\!P$Vs originated from the interference of even and odd waves.

\section{\boldmath Data-based analysis of FBA and FB-$C\!P$A in $B^\pm \to K^\pm K^\mp K^\pm$ \label{sec:AnalysisFBA}}
\begin{figure}[h!]
	\centering
	\includegraphics[width=1\textwidth]{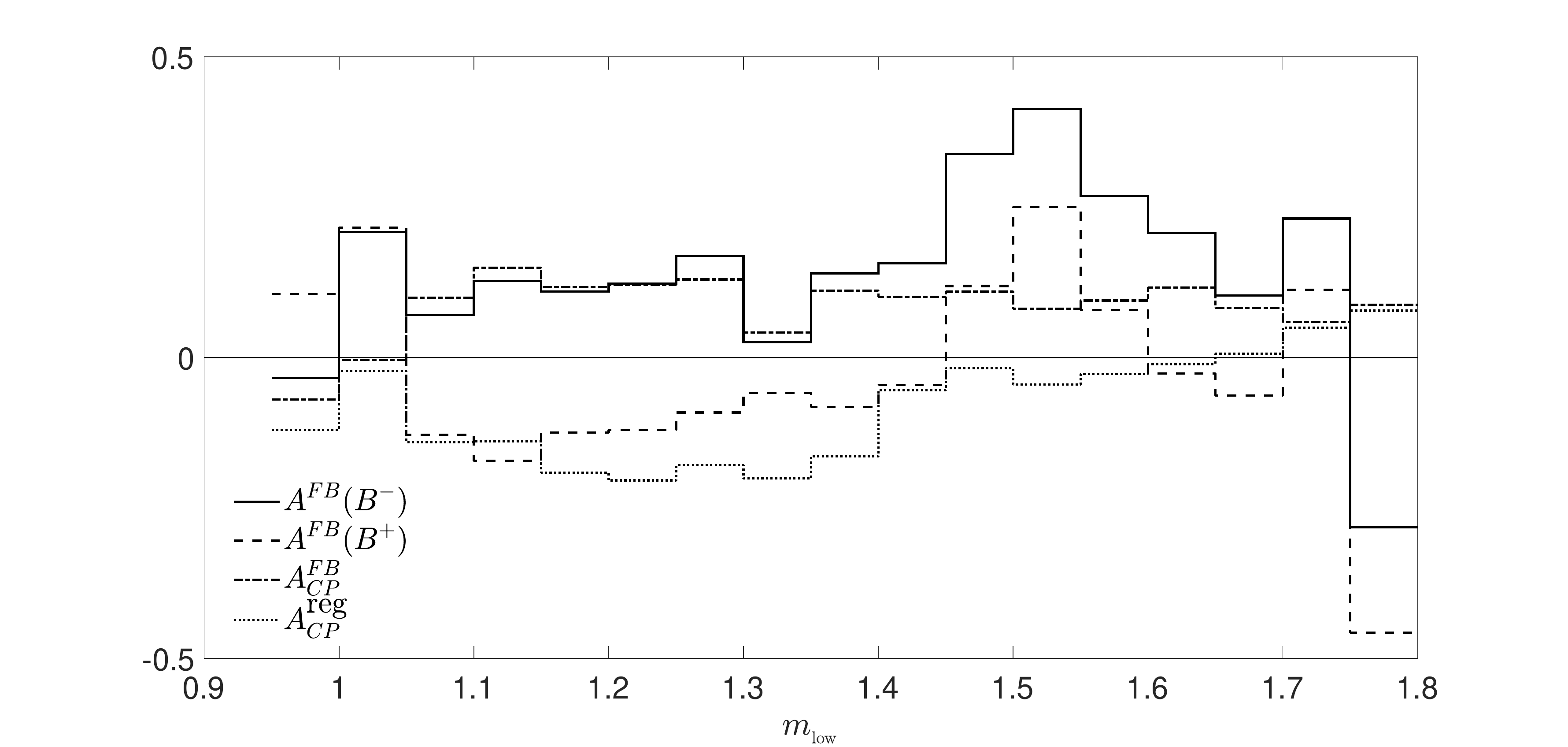}
	\caption{
Various observables extracted from the LHCb data in Ref. \cite{LHCb:2014mir}.
The solid and the dashed lines are the FBA for $B^-\to K^-K^+K^-$ and $B^+\to K^+K^-K^+$ respectively; the dash-dotted and the dotted lines are the FB-$C\!P$A $A_{C\!P}^{F\!B}$ and regional $C\!P$A $A_{CP}^{\text{reg}}$ respectively.
}\label{fig:datafigure}
\end{figure}

Thanks to the high statistics, the LHCb is able to investigate the $B$ meson decays --including the branching fractions and the $C\!P$ Asymmetries ($C\!P$As) of the three-body decays of $B$ meson-- in an unprecedented precision \cite{Chen:2021ftn,LHCb:2020xcz}.
A very detailed analysis has been carried out by LHCb for the decay process $B^\pm \to K^\pm K^\mp K^\pm$ mentioned in Sec. \ref{se:intro}, from which rich resonance structures and regional $C\!P$As can be clearly seen throughout the Dalitz plots \cite{LHCb:2014mir}.
Besides, signal yields projected in bins of the invariant mass of one of the $K^+ K^-$ pair were also investigated.
For each bin, the signal yield was divided into two parts according to whether $\cos\theta>0$ or $\cos\theta<0$, where $\theta$ was defined as the angle between the momenta of the two Kaons with the same-sign charge in the c.m. frame of the $K^+K^-$ pair with lower invariant mass, $m_{\text{low}}$ (also see FIG. \ref{fig:def-theta} for illustration).
Based on the data of FIG. 6 of the LHCb's paper \cite{LHCb:2014mir}, we obtain the measured FBAs, FB-$C\!P$A, and regional $C\!P$A for each bin of the decay process $B^\pm \to K^\pm K^\mp K^\pm$, which are presented in FIG. \ref{fig:datafigure}, respectively.
One interesting behaviour of the FB-$C\!P$A is that its value almost does not change for $m_{\text{low}}$ ranges from 1 Gev to 1.8 GeV, which deserves investigations from both the experimental and theoretical sides. However, since this is not what we focus on in this paper, we will simply skip this point from now on.
\begin{table}[h!]
  \begin{center}
        \caption{The event yields, FBAs, and FB-$C\!P$As of each bin for ${m_{\text{low}}}$ ranging between 1.30 and 1.65 GeV, where the uncertainties are statistical only. The FBAs and FB-$C\!P$As are obtained according to the definitions $A^{FB}_{i}\equiv {[N_i(\cos\theta>0)-N_i(\cos\theta<0)]}/{[N_i(\cos\theta>0)+N_i(\cos\theta<0)]}$, $\overline{A}^{FB}_{i}\equiv {[\overline{N}_i(\cos\theta>0)-\overline{N}_i(\cos\theta<0)]}/{[\overline{N}_i(\cos\theta>0)+\overline{N}_i(\cos\theta<0)]}$ and $A_{CP,i}^{FB}\equiv\frac{1}{2}(A^{FB}_{i}-\overline{A}^{FB}_{i})$, respectively.
      }
    \label{tab:IntHalfIntSR}
  \begin{tabular}{c||c|c|c||c|c|c||c}
    \hline\hline
    \multirow{2}{*}{bin (GeV)} &\multicolumn{3}{c||}{$B^-$} & \multicolumn{3}{c||}{$B^+$} & \multirow{2}{*}{$A_{CP,i}^{FB}$ (\%)} \\
    \cline{2-7}
     & $N_i(\cos\theta\!\!>\!\!0)$ & $N_i(\cos\theta\!\!<\!\!0)$  & $A_i^{FB}$ (\%) & $\overline{N}_i(\cos\theta\!\!>\!\!0)$ & $\overline{N}_i(\cos\theta\!\!<\!\!0)$  & $\overline{A}_i^{FB}$ (\%) &  \\
    \hline
    {1.30-1.35} &  $683\pm26$   & $649\pm25$    &  $2.6\pm2.7$   &  $942\pm31$  &  $1059\pm33$   & $-5.9\pm2.2$   &  $4.2\pm1.9$  \\
        \hline
    {1.35-1.40} &  $926\pm30$   & $698\pm26$    &  $14.0\pm2.5$   & $1038\pm32$   & $1223\pm35$    & $-8.2\pm2.1$   &  $11.1\pm1.7$  \\
        \hline
    {1.40-1.45} &  $1399\pm37$  & $1019\pm32$    &  $15.7\pm2.0$   &  $1286\pm36$  & $1408\pm38$    & $-4.5\pm1.9$   &  $10.1\pm1.4$  \\
        \hline
    {1.45-1.50} &  $1995\pm45$   & $986\pm31$    &  $33.9\pm1.7$   & $1728\pm42$   & $1360\pm37$    & $11.9\pm1.8$   & $11.0\pm1.2$   \\
        \hline
    {1.50-1.55} &  $1702\pm41$   & $706\pm27$    &  $41.4\pm1.9$   & $1646\pm41$   & $986\pm31$    & $25.1\pm1.9$   & $8.1\pm1.3$   \\
        \hline
    {1.55-1.60} &  $1351\pm37$   & $778\pm28$    &  $26.9\pm2.1$   & $1212\pm35$   & $1034\pm32$    & $7.9\pm2.1$   &  $9.5\pm1.4$   \\
        \hline
    {1.60-1.65} &  $1022\pm32$   & $671\pm26$    &  $20.7\pm2.4$   & $842\pm29$   &  $887\pm30$   &  $-2.6\pm2.4$   & $11.7\pm1.7$   \\
  \hline\hline
  \end{tabular}
  \end{center}
\end{table}

Another characterized feature of FIG. \ref{fig:datafigure} lies in the obvious large FBAs associated with some resonances sturcture when the invariant mass of the $K^+ K^-$ pair is around $1.5$ GeV for both the $C\!P$ conjugate processes $B^\pm \to K^\pm K^\mp K^\pm$, which indicates strongly the interference of odd- and even-partial waves according to the analysis in last section.
The FBA of this region is so large that it can even be clearly seen from the events distributions in the Dalitz plots.
In what follows, we will focus on this phase space region.
To be more specific, our analysis in this paper is perform only for ${m_{\text{low}}}$ ranges between 1.30 and 1.65 GeV in order to exclude the potential pollution of resonances such as $\phi(1020)$ and $f_0(1710)$.\footnote{Both $\phi(1020)$ and $f_0(1710)$ have little influence to the observed large FBA. For $\phi(1020)$, although it is one of the dominant resonances, but it is far away from the region of the observed large FBA and its width is narrow enough. Hence its effects to the observed large FBA is negligible even if it is a vector resonance. For $f_0(1710)$, on the other hand, although it is not far away for the region of the observed large FBA and has a relatively large decay width, but it can not be the reason for the observed large FBA as it is a spin-even scalar resonance.}
The corresponding event yields for $\cos\theta>0$ and $\cos\theta<0$, as well as the FBAs and FB-$C\!P$As, are presented in TABLE 1 for all bins of ${m_{\text{low}}}$ ranging between 1.30 and 1.65 GeV, where the uncertainties are statistical only.

\begin{figure}[h!]
	\centering
	\subfigure%[$B^{-}$]
    {	    \label{fityield.sub.1}
	 		\includegraphics[width=0.45\linewidth]{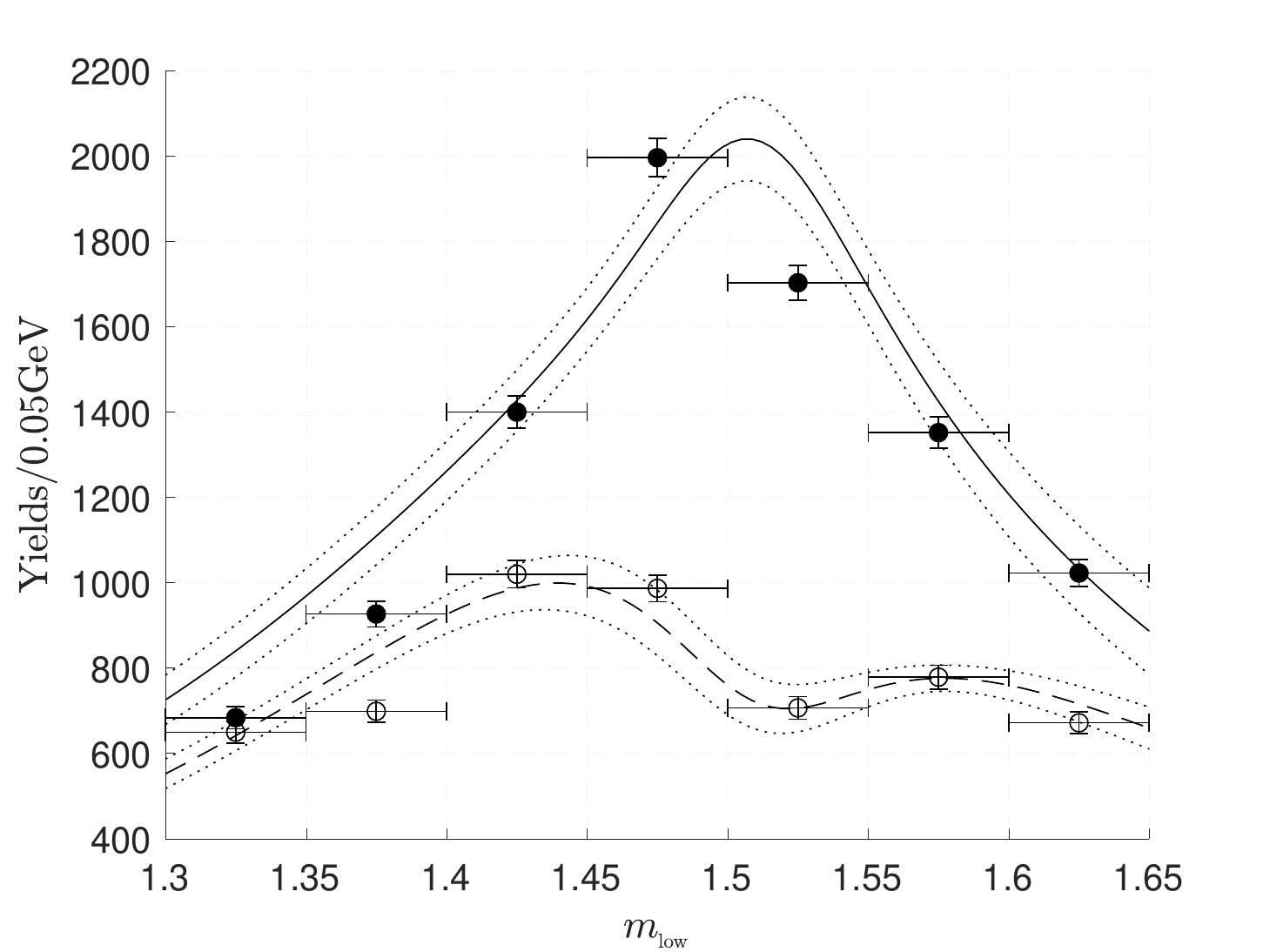}}
	 	%\quad %
 	\subfigure%[$B^{+}$]
    {	    \label{fityield.sub.2}
	 		\includegraphics[width=0.45\linewidth]{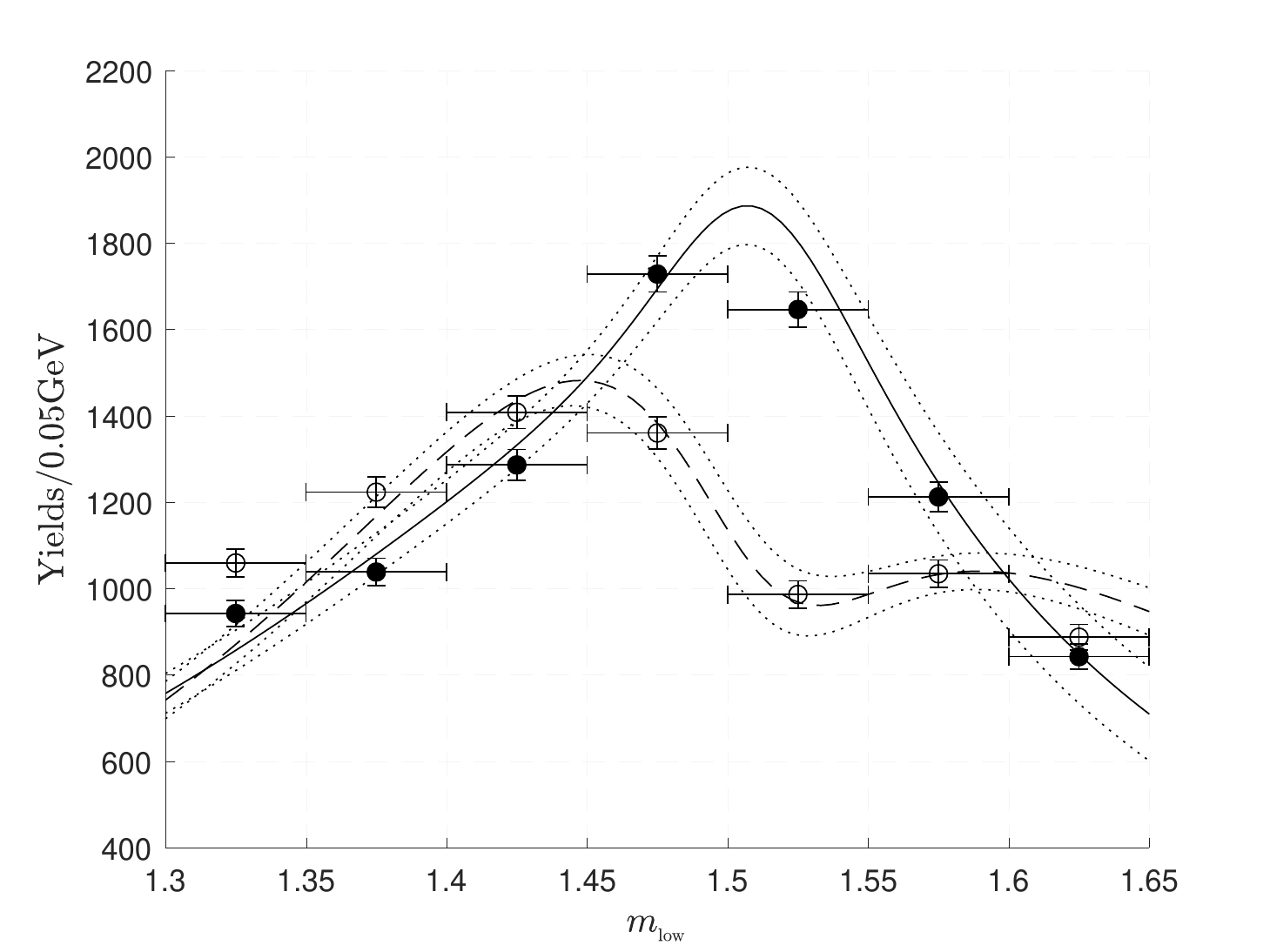}}
	        \caption{
The event yields from the data of LHCb (only the statistical error depicted) and its corresponding fitted curves of both the $C\!P$-conjugate decay channels $B^- \to K^- K^+ K^-$ (left) and $B^+ \to K^+ K^- K^+$ (right) projected in bins of the invariant mass of the $K^+ K^-$ ranging between 1.3 and 1.6 GeV, for $\cos\theta>0$ (solid circle for the data, solid line for the fitted curve) and $\cos\theta<0$ (hollow circle for the data, dashed line for the fitted curve). The dotted lines represent the range of the $1 \sigma$ confident-level fits.
}\label{fityield}
	\end{figure}

There are several resonances that could contribute to $B^\pm \to K^\pm K^\mp K^\pm$ in this region, including $f_0(1500)$, $\rho^0(1450)$, $X(1575)$, $f_2'(1525)$, etc., among which
the presence of $f_0(1500)$ and $f_2'(1525)$ has been reported by BaBar \cite{BaBar:2012iuj}.
After trying various fitting scenarios, we found that the best fit to the LHCb data of Ref. \cite{LHCb:2014mir} for ${m_{\text{low}}}$ ranges between 1.30 and 1.65 GeV constitutes of the resonances $f_0(1500)$ and $\rho^0(1450)$, plus a non-resonance $s$-wave.
The decay amplitude of $B^-\to K^- K^+ K^-$ can then be parameterized as
\begin{equation}%\label{}
  \mathcal{A}_{B^- \to K^- K^+ K^-}=\frac{c_1e^{i\delta_1}\cos\theta}{m^2_{\text{low}}-m_\rho^2+im_{\rho}\Gamma_\rho} +\frac{c_0e^{i\delta_0}}{m^2_{\text{low}}-m_f^2+im_{f}\Gamma_f}+\frac{c_{NS}e^{i\delta_{NS}}}{m_f\Gamma_f},
\end{equation}
where $c_l$ and $\delta_l$ ($l=0,1, NS$) are the corresponding amplitudes (excluding the corresponding propagators and the Legendre polynomials) and the relative phases, respectively, %which are the parameters to be fitted,
$m_{\rho/f}$ and $\Gamma_{\rho/f}$  are respectively the masse and the decay width of the resonance $\rho^0(1450)/f_0(1500)$.
The factor $1/m_f\Gamma_f$ in the last term is introduced to make sure that $c_{NS}$ has the same dimension with $c_0$ and $c_1$.
The amplitude of $B^+ \to K^+ K^- K^+$ can be obtained by replacing $c_l$ and $\delta_l$ by $\overline{c}_l$ and $\overline{\delta}_l$, respectively.
The event yield of each bin in FIG. \ref{fityield} is fitted according to
\begin{equation}%\label{}
  \mathcal{N}_{B\pm,i}(\cos\theta\gtrless 0)/0.05\text{GeV}= \int_{\cos\theta\gtrless 0}\Big[R\left|\mathcal{A}_{B^\pm \to K^- K^+ K^\pm}\right|^2\Big]_{m_{\text{low}}=\bar{m}_{\text{low},i}}  d \cos\theta,
\end{equation}
where $R=\sqrt{(m^2_{\text{low}}-4m_K^2) \left[m_B^2-(m_{\text{low}}-m_K)^2\right]\left[m_B^2-({m_{\text{low}}}+m_K)^2\right]}$ is the phase-space factor, and $\bar{m}_{\text{low},i}$ is the mean value of $m_{\text{low}}$ of bin $i$.
Note that we have absorbed all the factors which are irrelevant to the discussions of FBAs and FB-$C\!P$As into the amplitudes $c_l$.
Once this has been done, the amplitudes $c_l$'s become dimensionless.

\begin{table}[h!]
  \begin{center}
        \caption{The fitted values of the parameters $c_l$, $\delta_l$, $\overline{c}_l$, and $\overline{\delta}_l$, respectively. The phases $\delta_1$ and $\overline{\delta}_1$ are fixed to 0.
      }
    \label{tab:IntHalfIntSR}
  \begin{tabular}{c||c|c||c|c}
    \hline\hline
    resonance & $c_l$  & $\delta_l$ & $\overline{c}_l$  & $\overline{\delta}_l$ \\
    \hline
    $\rho^0(1450)$ & $30.7\pm0.5$ & 0 & $31.7\pm0.8$ & 0  \\
    \hline
    $f_0(1500)$ & $1.78\pm0.38$ &$-0.03\pm0.15$  & $2.12\pm0.33$ & $-0.27\pm0.13$   \\
    \hline
    non-res $s$-wave & $0.26\pm0.29$ &$2.20\pm0.80$  & $1.06\pm0.28$ & $2.15\pm0.19$   \\
    \hline\hline
  \end{tabular}
  \end{center}
\end{table}

The fitted curves are also presented in FIG. \ref{fityield} with inputs all taken from Ref. \cite{ParticleDataGroup:2020ssz}, while the numerical values of the fitted parameters are presented in TABLE \ref{tab:IntHalfIntSR}.
The goodness of the corresponding fits are 0.92 and 0.86 for $B^-\to K^\mp K^\pm K^\mp$ and $B^+\to K^\pm K^\mp K^\pm$ respectively, indicating that the data around $1.5$ GeV can be reasonably described by the resonances $\rho^0(1450)$ and $f_0(1500)$, with $\rho^0(1450)$ the dominant one.
This is in contradiction with the conclusion of BaBar in Ref. \cite{BaBar:2012iuj}, according to which the analysis showed no signal of the spin-1 resonance $\rho^0(1450)$.
With those fitted parameters, one can also obtain the $C\!P$ asymmetries of $B^\pm\to\rho^0(1450)K^\pm$, which is
  $A_{CP}(B^\pm\to\rho^0(1450)K^\pm)\equiv\frac{\left|c_1\right|^2-\left|\overline{c}_1\right|^2}{\left|c_1\right|^2+\left|\overline{c}_1\right|^2}=(-3.4\pm3.0)\%$.

\begin{figure}[h!]
	\centering
			 \includegraphics[width=.8\linewidth]{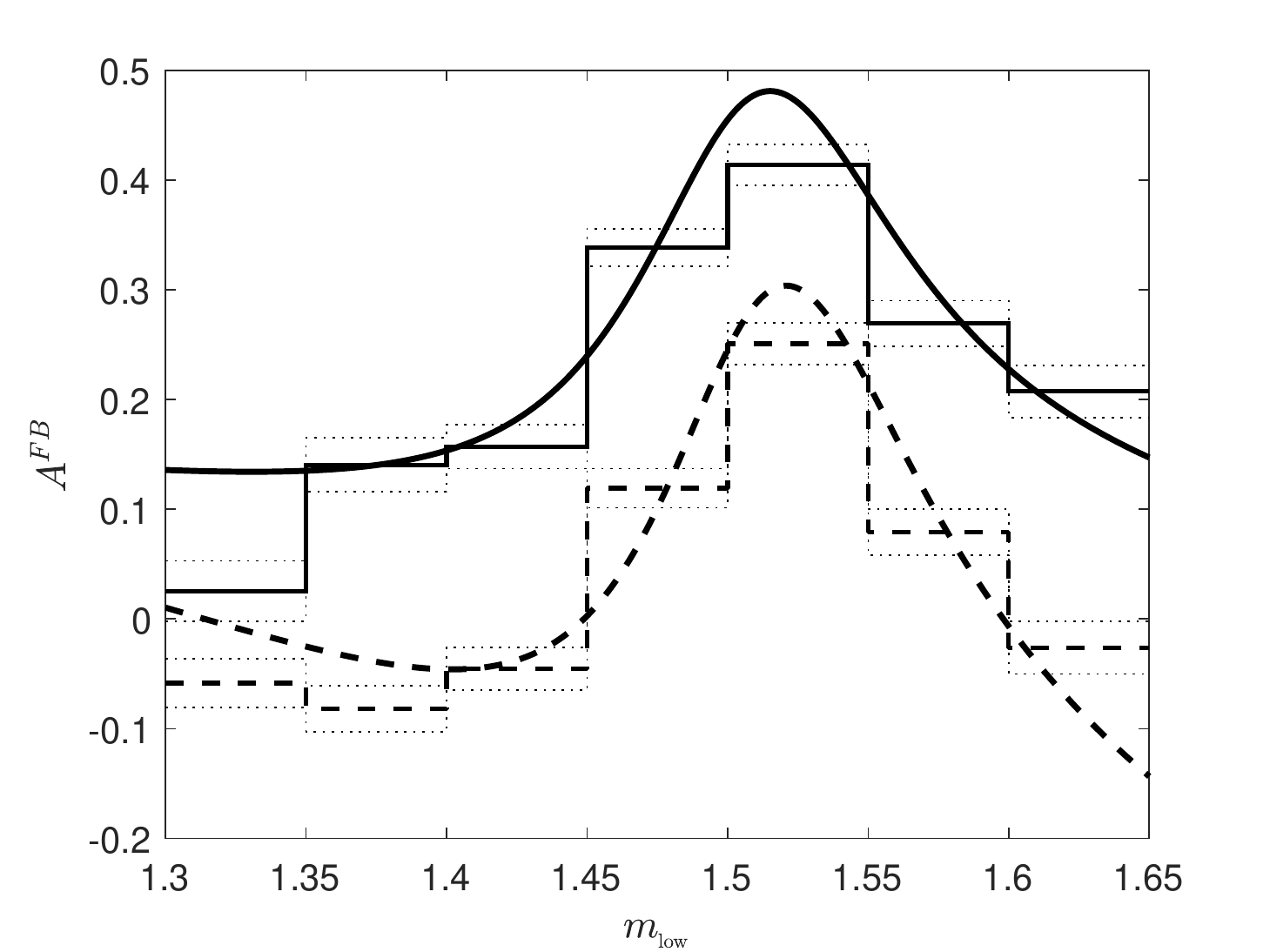}
			\caption{The best-fit FBAs of $B^{-} \to K^{-} K^{+} K^{-}$ (solid curve) and $B^{+} \to K^{+} K^{-} K^{+}$ (dashed curve) comparing with those extracted from the data of LHCb. The corresponding step lines are the FBAs extracted from the data. The dotted step lines represent the statistical uncertainties.}\label{fig:AFB}
		% \label{level}
\end{figure}

\begin{figure}[h!]
	\centering
			 \includegraphics[width=.8\linewidth]{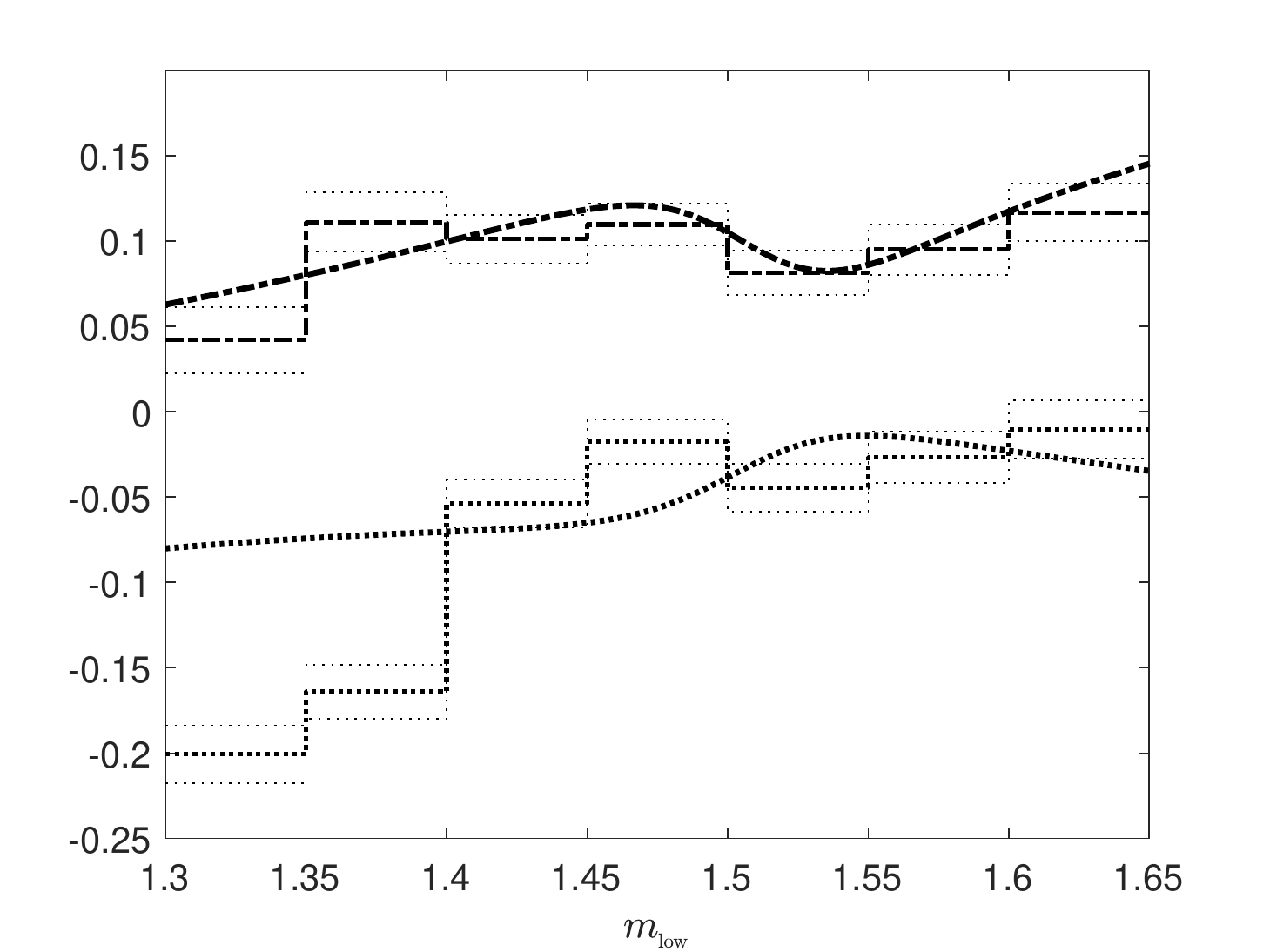}
			\caption{The best-fit FB-$C\!P$A (dash-dotted) and regional $C\!P$A (thick dotted) of $B^{\pm} \to K^{\pm} K^{\mp} K^{\pm}$. The corresponding step lines are the FBAs extracted from the data, while the curves are plotted with the input of the parameters taking values of the best fit. The thin dotted lines represent the statistical uncertanties.}\label{fig:CPV}
\end{figure}

With the central values of the fitted parameters, the FBAs of $B^\pm \to K^\pm K^\mp K^\pm$ are depicted in FIG. \ref{fig:AFB}, in along with comparisons with those obtained from the data.
One can see from this figure that the fitted FBAs fit with the data quite well, which is not a surprising result since this fit is in essence optimised according to the FBAs.
On the other hand, the fitted FB-$C\!P$As and regional $C\!P$As, which are presented in FIG. \ref{fig:CPV}, show less accordant with those from the data.
This is also understandable since both of the FB-$C\!P$As and the regional $C\!P$As represent ``fine structures'' of the decay $B^\pm \to K^\pm K^\mp K^\pm$ comparing with the FBAs.
Our analysis is too simple to describe such ``fine structures'' as the FB-$C\!P$As and the regional $C\!P$As well.
But be that as it may, these fitted curves in FIG. \ref{fig:CPV}, especially that of the FB-$C\!P$A, still show tendency that are in accordant with those from the data.

We also try other fitting scenarios, which are presented in TABLE \ref{tab:FitScenario}, along with the goodness of each scenario.
For example, we have try to fit the data by replacing $\rho^0(1450)$ by $X(1575)$, which has been observed by the BES collaboration in the channel $J/\psi \to K^+K^-\pi^0$ long time ago \cite{BES:2006kmo}, but the goodness of this fit is bad.
We have also try to fit the data by replacing $\rho^0(1450)$ by non-resonance $p$-wave. However, the fit is bad either, indicating that the large FBA for $m^2_{\text{low}}$ around 1.5 GeV can not either be explained by non-resonance $p$-wave contributions.
From TABLE \ref{tab:FitScenario} one can see that the fitting scenario which was presented in details above represents the best among all those in this table.

\begin{table}[h!]
  \begin{center}
    \caption{Various fitting scenarios that were performed. The goodness $\chi^2/\text{d.o.f.}$ for each fit are obtained with the inclusion of uncertainties from $m_{\text{low}}$, which were simply estimated as $0.05~\text{GeV}/2=0.025~\text{GeV}$ for each bin. The table is ordered according to the goodness of each fit.}
    \label{tab:FitScenario}
\begin{tabular}{c c c||c|c}
  \hline\hline
  \multicolumn{3}{c||}{fitting scenario}  & \multicolumn{2}{c} {$\chi^2/\text{d.o.f.}$} \\
  \hline
   $s$-wave & $p$-wave & $d$-wave &   $B^-$ & $B^+$\\
  \hline\hline
  $f_0(1500)$ & \multirow{2}{*}{$\rho^0(1450)$} &     & \multirow{2}{*}{0.92} & \multirow{2}{*}{0.86}\\
  non-res $s$-wave &                        & &                   &                     \\ \hline
   $f_0(1500)$ & $\rho^0(1450)$ & non-res $d$-wave  & 0.94 & 1.92 \\ \hline
  $f_0(1500)$ & $\rho^0(1450)$                  &  & 1.07 & 2.10 \\ \hline
  $f_0(1500)$ & $\rho^0(1450)$ & $f_2'(1525)$      & 1.10 & 2.57 \\ \hline
              & $\rho^0(1450)$ & $f_2'(1525)$      & 1.76 & 2.70 \\ \hline
  non-res $s$-wave & $\rho^0(1450)$ & $f_2'(1525)$ & 2.61 & 2.75 \\ \hline
  non-res $s$-wave & $\rho^0(1450)$ &              & 3.95 & 3.40 \\ \hline
  $f_0(1500)$ & non-res $p$-wave & $f_2'(1525)$  & 10.1 & 41.1 \\ \hline
  $f_0(1500)$ & $X(1575)$ &  $f_2'(1525)$        & 9.59 & 45.8 \\ \hline
  \hline
\end{tabular}

  \end{center}
\end{table}

\section{Summary and Conclusion}

A general analysis of the FBA and the FB-$C\!P$A were presented in this paper.
According to the analysis of this paper, the FBA as well as the FB-$C\!P$A are sensitive to the interfering effects of even-and odd-waves in three-body decays of  beauty and charmed mesons.
This makes them serve as good tools for the resonance structure analysis in the aforementioned decay processes.
We suggest our experimental colleagues to perform the measurements of the FBAs (as well as the FB-$C\!P$As) in three-body decay channels of beauty and charmed mesons.

Enlightened by the notably large FBAs embedded in the LHCb data in Ref. \cite{LHCb:2014mir}, we performed a data-based analysis of the FBAs and $C\!P$As of the decay $B^\pm \to K^\pm K^\mp K^\pm$.
We found that the large FBA observed when the invariant mass of the $K^+ K^-$ pair lies around 1.5 GeV can be interpreted as the interference of the amplitudes between the resonances $f_0(1500)$ with $\rho^0 (1450)$.
The analysis shows the existence of the decay channel $B^\pm\to \rho^0(1450) K^\pm$, with $C\!P$ asymmetry of $A_{CP}(B^\pm\to \rho^0(1450) K^\pm)=(-3.4\pm3.0)\%$.
This is in contradiction with the conclusion of BaBar in Ref. \cite{BaBar:2012iuj}, according to which the analysis showed no signal of the spin-1 resonance $\rho^0(1450)$.
We suggest our experimental colleagues to take a closer analysis on the decay channel $B^\pm \to K^\pm K^\mp K^\pm$.

\begin{acknowledgments}
We thank Wen-Bin Qian for useful discussion.
This work was supported by National Natural Science Foundation of China under Contracts Nos. 11705081 and 12192261.
\end{acknowledgments}

\bibliography{zzhbib}

\end{document}